# Dopant-Dopant Interactions in Beryllium doped Indium Gallium Arsenide: an Ab Initio Study

Vadym Kulish[1], Wenyuan Liu[2], Francis Benistant[3], Sergei Manzhos[1,a]

[1] Department of Mechanical Engineering, National University of Singapore, Block EA, 9 Engineering Drive 1, Singapore

[2] Division of Physics and Applied Physics, School of Physical and Mathematical Sciences, Nanyang Technological University, 21 Nanyang Link, Singapore 637371, Singapore

[3] GLOBALFOUNDRIES Singapore Pte Ltd., 60 Woodlands Industrial Park D Street 2, Singapore, 738406, Singapore

**ABSTRACT**

We present an *ab initio* study of dopant-dopant interactions in beryllium-doped InGaAs. We consider defect formation energies of various interstitial and substitutional defects and their combinations. We find that all substitutional-substitutional interactions can be neglected. On the other hand, interactions involving an interstitial defect are significant. Specially, interstitial Be is stabilized by about 0.9/1.0 eV in the presence of one/two $BeG_a$ substitutionals. Ga interstitial is also substantially stabilized by Be interstitials. Two Be interstitials can form a metastable Be-Be-Ga complex with a dissociation energy of 0.26 eV/Be. Therefore, interstitial defects and defect-defect interactions should be considered in accurate models of Be doped InGaAs. We suggest that In and Ga should be treated as separate atoms and not lumped into a single effective group III element, as has been done before. We identified dopant-centred states which indicate the presence of other charge states at finite temperatures, specifically, the presence of $Be_{int}^{+1}$ (as opposed to $Be_{int}^{+2}$ at 0K).

[a] Address all correspondence to this author: mpemanzh@nus.edu.sg (S. Manzhos)

## I. INTRODUCTION

The ternary compound InGaAs has received significant research interest due to its considerably higher electron mobility compared to silicon and a lattice constant that matches with that of InP. These factors make it a promising candidate for future complementary metal-oxide-semiconductor (CMOS) devices, specifically, for the cutting-edge technologies e.g. 5 nm node and below.[1] Because of high activation ratio and well-developed and controllable doping methods, beryllium is considered as an attractive and important *p*-type dopant for InGaAs. Therefore, much effort on experiments and simulations has been devoted to investigate and understand the Be doping mechanism and diffusion behaviour. Be diffusion in InGaAs is abnormally fast: the diffusivity is about five orders of magnitude larger than Be diffusion in GaAs at the same temperature,[2,3] and the mechanism of it has been under debate. Specifically, kick-out and Frank-Turnbull mechanisms were proposed,[3,4] but a single one of them has not been able to explain experimental diffusion profiles obtained at different temperatures. A number of approximations are usually made in theoretical studies of the Be diffusion mechanism, such as the neglect of the difference between the two group III elements as well as neglect of the effect of As. These approximations can much simplify simulation; however, they were not supported by in-depth analysis and specifically ab initio analysis. For example, in previous continuum or Monte Carlo studies, parameters entering the model such as charge states and reaction energies / diffusion barriers for elementary reactions – on which the diffusion rate critically depends - were postulated or fitted rather than derived from first principles.[5,6] The experiments on Be diffusion are still relatively limited, which means that by tuning parameters and postulating different diffusion mechanisms one can achieve a good fit to diffusion profiles obtained in specific experimental conditions[3,4,7], however, such fitting is not very meaningful because of the number of approximation.

In a recent multiscale study combining *ab initio* calculations with continuum and kinetic Monte Carlo simulations[8], we provided *ab initio* reaction energies and diffusion barriers which suggested that the kick-out mechanism is preferred to the Frank-Turnbull mechanism. The *ab initio* results also suggested that In and Ga may have rather different roles in Be diffusion, for example, kicking out of Ga is an exothermic reaction while kicking out of In is endothermic. Therefore In and Ga should not be lumped into a single effective group III element. Some reaction energies involving As are comparable with reactions involving Ga and In, which suggests that As may also play a role in Be diffusion process[8]. *Ab initio* simulations also provided 0 K atomic charge states as well as estimates of finite-temperature charge state distributions, based on the analysis of the electronic structure (densities of states). *Ab initio* based charge estimates provide a more solid basis for diffusion simulation compared to when the charges were assigned (in previous models) by fitting to experiment and or based on intuition. The continuum and Monte Carlo simulations confirmed that temperature-dependent charge states in agreement with *ab initio* results provide a good match to experimental diffusion profiles measured at different temperatures. Different charge states and therefore different reaction will dominate at different temperature. They also confirmed the preference for the kick-out mechanism.

The *ab initio* model used in Ref. [8] made a number of simplifying assumptions: in particular, isolated interstitial and substitutional defects were considered. Here, we are interested in the effect of interaction of different defects on properties determining Be diffusion such as defect formation energies at 0 K as well as possible temperature-dependent charge states. Defect interactions are potentially important, as formation of stabilized defect pairs may lead to concerted diffusion observed in other systems.[9] It is also highly likely to lead to changes in both 0 K atomic charges and in densities of states which may lead to different finite-temperature charge distributions. Specifically, interstitial Be is an *n*-dopant while substitutional Be is a *p*-dopant, and significant interactions are expected between them which in other systems have been shown to lead to substantial stabilization.[10, 11] Is this also the case in Be-doped InGaAs? Specifically, in Ref. [11] we showed that interactions between substitutional (*p*-dopant) and interstitial (*n*-dopant) Mg (as well as Li and Na) in Ge substantially affect defect formation energies and qualitatively change the electronic structure. In Refs. [10, 11] we also showed that *p*-dopants Al in Si and Ga in Ge substantially stabilize interstitial Mg (as well as Li and Na). One also expects significant interactions between interstitial defects; for example, mono- and di-valent interstitial defects in group IV monoelemental materials such as Si or Sn repel each other near-Coulombically.[12, 13] For example, in Refs. [11, 14, 15] we and others showed that interstitial Mg as well as Li, Na in Si and Sn repel each other. Does this picture hold for Be in InGaAs?

Previously, we showed that tri-valent substitutional *p*-dopants in (four-valent) Si repel each other.[10] For example, in Refs. [10, 11] this was shown that substitutional *p*-dopants (Al and Ga) in Si and Ge. What is the nature of interactions between substitutional Be atoms in InGaAs? Some self-interstitials in InGaAs are energetically allowed;[3] how will they affect interstitial or substitutional Be dopants? These are the questions that we address in this paper.

Theoretical studies of doped multi-elemental semiconductors often rely on estimates of chemical potentials of components to compute defect formation energies[16, 17]. This is because, contrary to doped monoelemental semiconductors, it is not easy to assign an energy of a single atom in a multicomponent solid, in the case of substitutional defects. The use of chemical potentials implies an open system, and the resulting defect formation energies are not uniquely defined, as chemical potentials can assume a range of values for various so-called component-rich or component-poor conditions. This is in contrast to doping of monoelemental semiconductors, where one can uniquely compute defect formation energies, because all reference states are well-defined; specifically, the defect formation energy of a substitutional defect only requires the knowledge of the formation energy of the host and of the bulk state of the dopant. Well-studied examples are those of alkali *and alkali earth* atoms in Si, Ge, Sn mentioned above[10-15, 18-26]. This has the advantage of the calculation being fully *ab initio* with uniquely defined formation energies. One novelty proposed here is the use of a model we have previously introduced in Ref. [27] that allows considering Be-doped InGaAs as a closed system and uniquely define formation energies without recourse to chemical potentials. This allows us to compute fully *ab initio* and compare different types of defects (substitutional, interstitial, and their combinations).

Many *ab initio* studies of doped semiconductors postulate charge states of defects rather than derive them *ab initio*[16, 17, 28, 29]. Charge states can be selected to fit proposed diffusion models, as has been done specifically for the case of Be-doped InGaAs [7, 30, 31] or to satisfy chemical intuition. Such calculations are typically done in DFT (density functional theory) using charged cells with background compensating charge (to avoid divergence of total energy)[17, 28, 32-34]. This amounts to forcing the DFT compute what charge state one wants there to be in the first place instead of using the predictive power of *ab initio* calculations. For example, in Ref. [3], the authors "*suggest* the case of neutral beryllium interstitials" (emphasis is ours). This is in spite of a well-known propensity of interstitial alkali earth atoms to donate both their valence electrons to the conduction band of the semiconductor host [10, 11, 13, 35]; the donation of two electrons to the conduction band was also explicitly computed for interstitial Be in InGaAs in Ref. [8]. In Ref. [8], we proposed an *ab initio* approach to charge states whereby DFT is used to compute the charge states at 0 K (on a neutral cell, in agreement with the fact that bulk material is neutral), and electronic structure analysis is used to identify defect-centred states which could potentially be occupied at finite temperature, giving rise to a temperature-dependent distribution of charge states. Such *ab initio* derived charges were validated in continuous and Kinetic Monte Carlo models of Be diffusion which allowed to reproduce experimental diffusion profiles over a range of temperatures [8]. Here, we also use this approach. The non-integer charge states computed by the analysis of the change density (Bader charges) are assigned integer values based on qualitative analysis of the electronic structure.

In this paper, we therefore present a fully *ab initio* investigation of most energetically favoured interstitial and substitutional defects in Be-doped InGaAs and of interactions between them, as well as their charge states. Kinetic properties on the other hand are not the considered here. The paper is organized as follows: Section II details the methods used and computational parameters, Section III presents the resulting energies and charge states of single and double defects of different types as well as their effect on the electronic structure (partial densities of states), and Section IV concludes.

## II. METHODS

The calculations were performed using density functional theory (DFT) with the generalized gradient approximation (GGA) and the Perdew-Burke-Eznerhof functional (PBE)[36] as implemented in the Vienna Ab Initio Simulation Package (VASP) 5.3 package.[37] The core electrons were treated within the projector augmented wave (PAW) method.[34, 38] The following valence electron configurations were used: arsenic (As) $4s^24p^3$, gallium (Ga) $4s^24p^1$, indium (In) $5s^25p^1$, and beryllium (Be) $2s^22p^0$. The plane-wave basis set cut-off energy was set at 400 eV which provided converged values. The Brillouin zone was samples with Monkhorst-Pack k-point meshes.[39] Monkhorst-Pack k-point meshes of 4x4x4 and 5x5x5 were used for structure optimization and density-of-state (DOS) calculations, respectively. Contributions to the DOS from different types of atoms (partial densities of states, PDOS) were analysed. The optimized structures were obtained by relaxing all atomic using the conjugate gradient algorithm until all forces were smaller than 0.02 eV Å$^{-1}$. To ensure that the simulation cell is of size amenable to the calculations, we used the stoichiometry In$_{0.5}$Ga$_{0.5}$As$_1$ (abbreviated in the following as InGaAs) and simulation cell size of about 12.0x12.0x11.9 Å, as was done in previous works.[8, 32, 33, 40] This cell size is sufficient in size to minimize interactions between periodic images; the cell vectors were kept fixed during defect optimization. The Bader charges on atoms in pure InGaAs are: Ga: +0.65, In: +0.67, and As: -0.64 |$e$|.

The defect formation energies are computed as

$$E_f = \frac{E_d - E_{ideal} - nE_{int} + mE_{sub}}{n} \tag{1}$$

where $n$ is the number of inserted atoms (either at substitutional or interstitial position) and $E_{int}$ is the energy of the inserted atom in its aggregate state (bulk Be, In, Ga, As), $m$ is the number of removed host atoms in the case of substitutional defects (in defects considered here, $m < n$) and $E_{sub}$ is the energy of the removed atom, i.e. the energy of In, Ga, or As in InGaAs ($E$(In), $E$(Ga), and $E$(As) described below), $E_d$ is the energy of the defected simulation cell (with $n$ inserted and $m$ removed atoms), and $E_{ideal}$ is the energy of the non-defected cell (ideal InGaAs). Note that $E_f$ is defined here per defect. While for single and double interstitial defects, the application of Eq. (1) is straightforward, this is not so for substitutional defects in a multicomponent solid. When analysing defects in multicomponent solids, one typically estimates the chemical potential of components, and the end result is that typically the defect formation energies are ambiguously given as a function of chemical potential or Fermi level.[17, 33, 41, 42] One ends up with a wide range of $E_f$ values for various so-called component-poor or component-rich conditions. This corresponds to modelling an open system and makes the computed $E_d$ rather non-*ab initio* and certainly not unique. In Ref. [27], we introduced a way of computing $E_f$ from only structures and energies of compounds, similarly to what is typically done for defect in monoelemental materials[10-15, 18-26]. This is achieved by assigning a fraction $f$ of formation energy $E_{form}$ of InGaAs (computed vs bulk In, Ga, As) is assigned to each type of constituent atoms (In, Ga, As) as

$$E_{form}(\text{InGaAs}) = E_{form}(\text{InGaAs}) \times (f_{In} + f_{Ga} + f_{As}) = (E(\text{In}) + E(\text{Ga}) + 2E(\text{As}))/4 \tag{2}$$

where $E(\text{In})$, $E(\text{Ga})$, and $E(\text{As})$ are chosen to satisfy formation energies of In, Ga, and As containing materials of several stoichiometries:

$$E_{form}(\text{In}_x\text{Ga}_y\text{As}_z) = (xE(\text{In})+yE(\text{Ga})+zE(\text{As}))/(x+y+z) \qquad (3)$$

Here, zero values of one of $x, y, z$ are possible, i.e. binary and ternary compounds are considered. We include $\text{InGaAs}_2$, $\text{In}_3\text{Ga}$, GaAs, InAs, $\text{In}_3\text{Ga}_5\text{As}_8$, $\text{In}_5\text{Ga}_3\text{As}_8$, $\text{In}_{16}\text{Ga}_{17}\text{As}_{31}$, $\text{In}_{16}\text{Ga}_{15}\text{As}_{33}$[27] as materials whose formation energies we aim to reproduce with Eq. (3), resulting in eight sets of $(x, y, z)$ and in a rectangular matrix equation

$$Ac = B \qquad (4)$$

where

$$A = \begin{pmatrix} 1/4 & 1/4 & 2/4 \\ 3/4 & 1/4 & 0 \\ & \cdots & \\ 16/64 & 15/64 & 33/64 \end{pmatrix}, \quad B = \begin{pmatrix} E_{form}(\text{InGaAs}) \\ E_{form}(\text{In}_3\text{Ga}) \\ \cdots \\ E_{form}(\text{In}_{16}\text{Ga}_{15}\text{As}_{33}) \end{pmatrix}, \quad c = \begin{pmatrix} E(\text{In}) \\ E(\text{Ga}) \\ E(\text{As}) \end{pmatrix} \qquad (5)$$

where the matrix $A$ contains the stoichiometric coefficients (in which the sum of values in each rows equals to one), vector $B$ contains the formation energies we aim to match by fitting $E(\text{In})$, $E(\text{Ga})$, and $E(\text{As})$, and the vector $c$ contains the values of $E(\text{In})$, $E(\text{Ga})$, and $E(\text{As})$. This equation in general will not have an exact solution but can be solved in the least-squares sense, $c = pinv(A) \cdot B$, where $pinv$ stands for the pseudoinverse.[43] This gives

$$\begin{pmatrix} E(\text{In}) \\ E(\text{Ga}) \\ E(\text{As}) \end{pmatrix} = \begin{pmatrix} 0.108 \\ -0.120 \\ -0.468 \end{pmatrix} \text{eV} \qquad (6)$$

The residual $R = Ac - B$ is then a measure of the accuracy of this approximation. Based on the eight compounds listed above, the errors in formation energies of all structures (computed from the residual $R$) are on the order of 0.05 eV/atom[27] and are acceptable for the purpose of this work. In Ref. [27] we showed that the errors in formation energies are not very sensitive to the exact composition of the set of structures used to fit $E(\text{In})$, $E(\text{Ga})$, and $E(\text{As})$; for example, a fit using only four reference structures (InGaAs, $\text{In}_3\text{Ga}$, GaAs, and InAs) resulted in errors in the formation energy of about 0.04 eV (mean absolute error) of the *eight* structures and values of $E(\text{In})$, $E(\text{Ga})$, and $E(\text{As})$ different by less than 0.05 eV from those listed above (Eq. (6)).[27] This essentially one-body approximation, which does not carry structural (or atomic coordination) information, is nonetheless useful for the specific purpose of this work, i.e. assigning fractions of the formation energy of a ternary compound to its constituent atoms. Then the approximation for the energy of a substituted atom of type $X$ ($X$ = In, Ga, As), needed to computed $E_f$ for substitutional defects ($E_{sub}$ in Eq. (1)), would be simply

$$E_{sub}(X) = E(X_{bulk}) + f_X \times E_{form}(\text{InGaAs}) = E(X_{bulk}) + E(X) \quad (7)$$

where $E(X_{bulk})$ is the energy per atom of atom $X$ in its reference (bulk) state and $E(X)$ is as in Eq. (6). This approach, which effectively allows us to treat doped BeGaAs as a closed system (as opposed to an open system requiring the use of chemical potentials) in a similar way to that in which dopants in monoelemental hosts are typically treated[10-15, 18-26], permits assigning unique, purely *ab initio* values to defect formation energies.

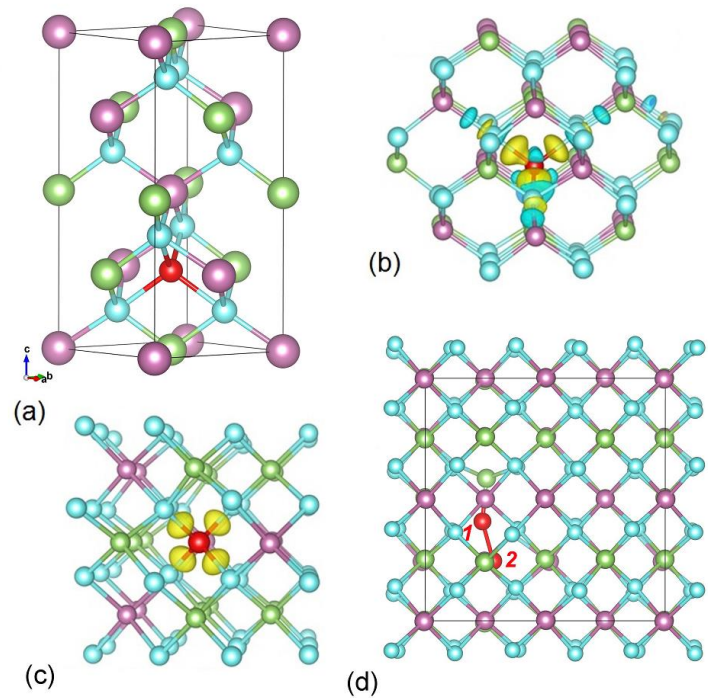

FIG. 1. (a) Unit cell of InGaAs with a single dopant at the interstitial site. The In, Ga, As and interstitial atoms are shown in violet, green, blue and red colours, respectively. (b, c) Isosurfaces of charge density difference for InGaAs with (b) Be interstitial and (c) Be substitutional. The yellow/blue colours show charge accumulation/ depletion, respectively. (d) The defect structure

(Be-Ga dumbbell) formed when two Be interstitials are near. Lines show the limits of the simulation cell.

## III. RESULTS AND DISCUSSION

### A. Single interstitial and substitutional defects

The unit cell of InGaAs is shown in Fig. 1(a); the simulation cell is 2×2×1 unit cells. We are thus using a "regular" alloy as a model. We tested the effect of randomness in Ref. [8] and found that they were small; specifically, no significant changes in the band structure are caused by randomness and changes in reaction energies were small when exchanging the positions of In and Ga. Below we also show that In and Ga substitutional defects on Ga and In, respectively, have defect formation energies relatively close to zero and do not perceptibly modify the density of states (contrary to single or double defects involving interstitials considered below). Therefore the conclusions are not expected to be affected by the randomness of the alloys, although a detailed study of these effects is still outstanding but us not the subject of the present paper. The sites of the substitutional defects correspond to lattice In, Ga, and As sites; the interstitial site is explicitly indicated in Fig. 1(a). At the interstitial site, the inserted atoms occupy the tetrahedral position. The defect formation energies of single interstitial and substitutional defects are given in Table I. These include Be interstitial ($Be_{int}$), Be substitutional defects at In, Ga, and As sites ($Be_{In}$, $Be_{Ga}$, and $Be_{As}$, respectively), In interstitial ($In_{int}$) and In substitutional defects at Ga and As sites ($In_{Ga}$, and $In_{As}$, respectively), Ga interstitial ($Ga_{int}$) and Ga substitutional defects at In and As sites ($Ga_{In}$, and $Ga_{As}$, respectively), as well as As interstitial ($As_{int}$) and As substitutional defects at In and Ga sites ($As_{In}$, and $As_{Ga}$, respectively). It follows from Table I that Be strongly prefers substitutional position at indium and gallium sites, with $Be_{Ga}$ ($E_f = 0.18$ eV) preferred to $Be_{In}$ by about 0.24 eV, which highlights different roles played by different group III atoms. The $Be_{As}$ substitutional, on the other hands, can be ignored in view of its high energy ($E_f = 2.53$ eV). The interstitial Be possesses rather high energy ($E_f = 1.95$ eV) but is considered in the following, as long-range Be diffusion must involve a form of Be interstitial. We also show below that $Be_{int}$ (as well as $Ga_{int}$) is stabilized in the presence of substitutional defects.

**TABLE I.** Calculated formation energies of single defects in Be-doped InGaAs. Here and elsewhere, subscript "int" is used for interstitial defects and subscripts "In", "Ga", "As" for corresponding substitutional defects.

| Defect | Formation energy (eV) | Charge on dopant, $|e|$ |
|---|---|---|
| $Be_{int}$ | 1.952 | 1.35 |
| $Be_{In}$ | 0.414 | 1.41 |
| $Be_{Ga}$ | 0.176 | 1.42 |
| $Be_{As}$ | 2.532 | 0.88 |
| $In_{int}$ | 2.704 | 0.43 |
| $In_{Ga}$ | 0.012 | 0.70 |
| $In_{As}$ | 1.636 | 0.05 |
| $Ga_{int}$ | 1.810 | 0.36 |
| $Ga_{In}$ | -0.137 | 0.60 |
| $Ga_{As}$ | 1.272 | −0.05 |
| $As_{int}$ | 2.715 | -0.37 |
| $As_{In}$ | 1.537 | +0.21 |
| $As_{Ga}$ | 1.565 | +0.23 |

A $Ga_{In}$ is energetically favourable ($E_f$ = -0.14 eV i.e. negative) and is preferred to $Ga_{As}$ by a significant 1.4 eV. The Ga interstitial is also a high-energy defect with the defect formation energy ($E_f$ = 1.81 eV) of a similar magnitude to that of $Be_{int}$. $In_{Ga}$ substitutional is computed to be energetically relatively favourable with $E_f$ close to 0, while $In_{As}$ and $In_{int}$ are high-energy ($E_f$ = 1.64 eV and 2.70 eV, respectively). All As interstitial and substitutional defects are high-energy ($E_f$ = 2.72 eV, 1.54 eV, and 1.57 eV for $As_{int}$, $As_{In}$, and $As_{Ga}$, respectively) and are not further considered.

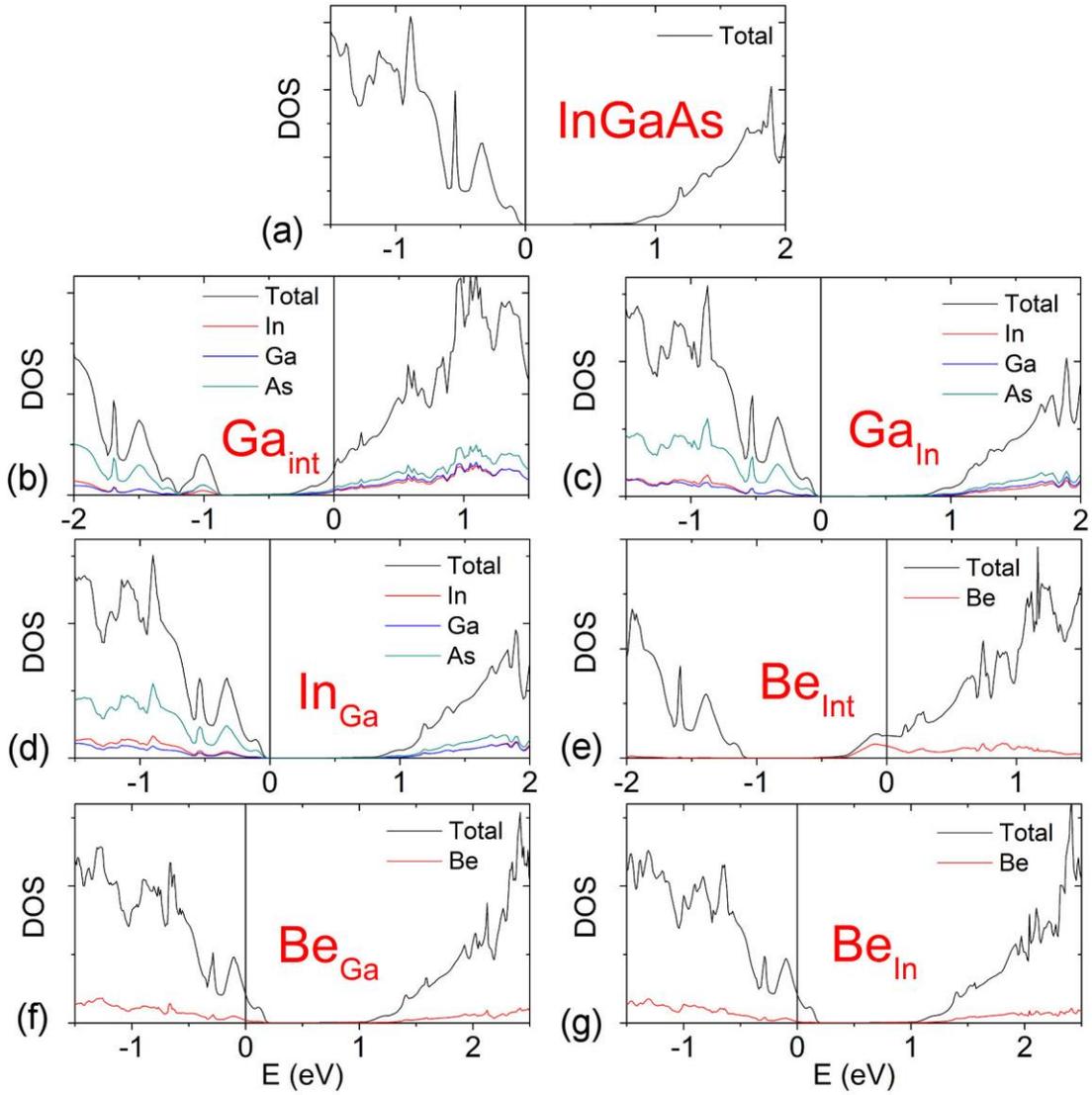

FIG. 2. Densities of states (DOS) of InGaAs: (a) pure, (b) with Ga interstitial, (c) with Ga substitutional, (d) with In substitutional, (e) with Be interstitial, and (f-g) with Be substitutional defects. Zero energy is set to the Fermi level. Be contribution to the DOS is magnified by 15.

The formation of defects leads to structural distortions in the host lattice. We found that the first nearest-neighbouring As atoms around the Ga and In interstitial defects move outward (away from the interstitial) after structural relaxation. This results in the stretching of related Ga-As and In-As bonds by about 0.09 and 0.12 Å, respectively. These values are quite significant, contributing to the energetic instability of In and Ga interstitials (see Table I). In contrast, formation of Ga and In substitutionals leads to only minor changes in Ga - As and In - As bond lengths (0.02 and 0.04 Å, respectively). Much reduced atomic relaxations in the case of substitutional defects help explain their low formation energies (Table 1) and energetic preference over interstitials. A similar argument holds for Be defects. In all interstitial and substitutional positions, Be maintains a tetrahedral coordination, being surrounded by four As atoms. After structural relaxation, the neighbouring As atoms are displaced from their equilibrium positions by 0.03, 0.04 and 0.09 Å due to the accommodation of $Be_{Ga}$, $Be_{In}$ and $Be_{int}$ defects, respectively. These values follow the same trend as the trend of the formation energies of the respective defects ($E_f(Be_{Ga}) < E_f(Be_{In}) < E_f(Be_{int})$, see Table I).

The electronic structure modifications induced by individual defects can be seen in Fig. 2 where the DOS are shown for pure as well as defected systems for lowest-energy defects. The corresponding Bader charges on the defects are listed in Table I. High energy defects are not shown. It is seen from Fig. 2 that Be interstitial acts an $n$ dopant while Be substitutionals on In and Ga act as $p$ - dopants. In the presence of defects, the DOS remains non- spin polarized and the total magnetic moment of the system is zero. The electronic structure is not perceptibly modified by $In_{Ga}$ and $Ga_{In}$ and the charge states of atoms are also not significantly modified, as expected. The interstitial Be donates both of its valence electrons to the conduction band; this is reflected in a Bader charge about +1.4 $|e|$ which we assign to an integer charge state of Be of +2 (integer charge states are useful for continuum and KMC models) due to donation of two electrons. In the following subsection, this assignment is justified based on electronic structure (DOS) analysis. The substitutional $Be_{In}$ and $Be_{Ga}$, which also have Bader charges of about +1.4 $|e|$, donate their valence electrons to the valence band; the Fermi level shifts into the valence band corresponding to the introduction of hole states. Below, we show that one substitutional Be atom creates one hole state (i.e. one empty electronic state at the top of the valence band). The "donation" in this case leads to bond formation with neighbouring atoms, as is confirmed in charge density difference analysis shown in Fig. 1(b,c). The charge-density difference ($\Delta\rho$) is calculated as:

$$\Delta\rho = \rho(Be+host) - \rho(host) - \rho(Be)$$

where $\rho$(Be+host), $\rho$(host) and $\rho$(Be) are total charge densities of Be - doped InGaAs, pristine InGaAs (the host) and an isolated Be atom. Note that all Be, In, Ga and As atoms are in the exact same positions as they occupy in the Be - doped InGaAs system. We can notice that charge is accumulated in the middle of Be - As bonds in Fig. 1(b,c). Notice the depletion of the electron density on the Be interstitial atom in Fig. 1b, indicating that there is charge transfer from Be to surrounding As atoms. This is further confirmed by the Bader charge analysis. Specifically, the Bader charges on As atoms in the vicinity of the Be interstitial change significantly from -0.64 |$e$| to -1.01 |$e$| as a result of Be insertion. The direction of electron transfer can be rationalized from the electronegativities of the elements (1.57 and 2.18 on the Pauling scale for Be and As, respectively). In the case of an interstitial Be, there is a peak in the partial DOS for Be (the red curve in Fig. 2) about 0.044 eV above the Fermi level, i.e. there is an empty (at 0 K) Be centred state which can be occupied with the presence of thermal energy. This implies that at a finite temperature, Be has a chance of having a charge state of +1. Based on thermodynamics (using the Fermi-Dirac occupation function), 18 % of Be interstitials would be in the +1 state at 300 K. However, that state is not fully Be centred, so a smaller fraction of $Be_{int}^{+1}$ is expected in reality. No such low-lying (above the Fermi energy) states are identified for Be substitutionals.

### B. Defect-defect interactions

We now investigate interactions between those defects that are expected to play a role in Be diffusion,[8] as identified in the previous section based in particular on the defect formation energies. These include substitutional Be, In, and Ga as well as interstitial Be and Ga, which, although higher-energy then interstitial Be and Ga, can be significantly stabilized as shown below; Be interstitials are also expected to be involved in diffusion. We therefore first study the interaction between two Be interstitials. We placed two such interstitials at different distances; the resulting defect formation energies per $Be_{int}$ are shown in Fig. 3(a,b) together with their Bader charges. At large separation, the defect formation energy per defect approaches that of a single defect (Table I), as expected. As Be-Be distance decreases, the defect formation energy increases; this is expected given significant positive charges on Be which lead to coulombic repulsion.

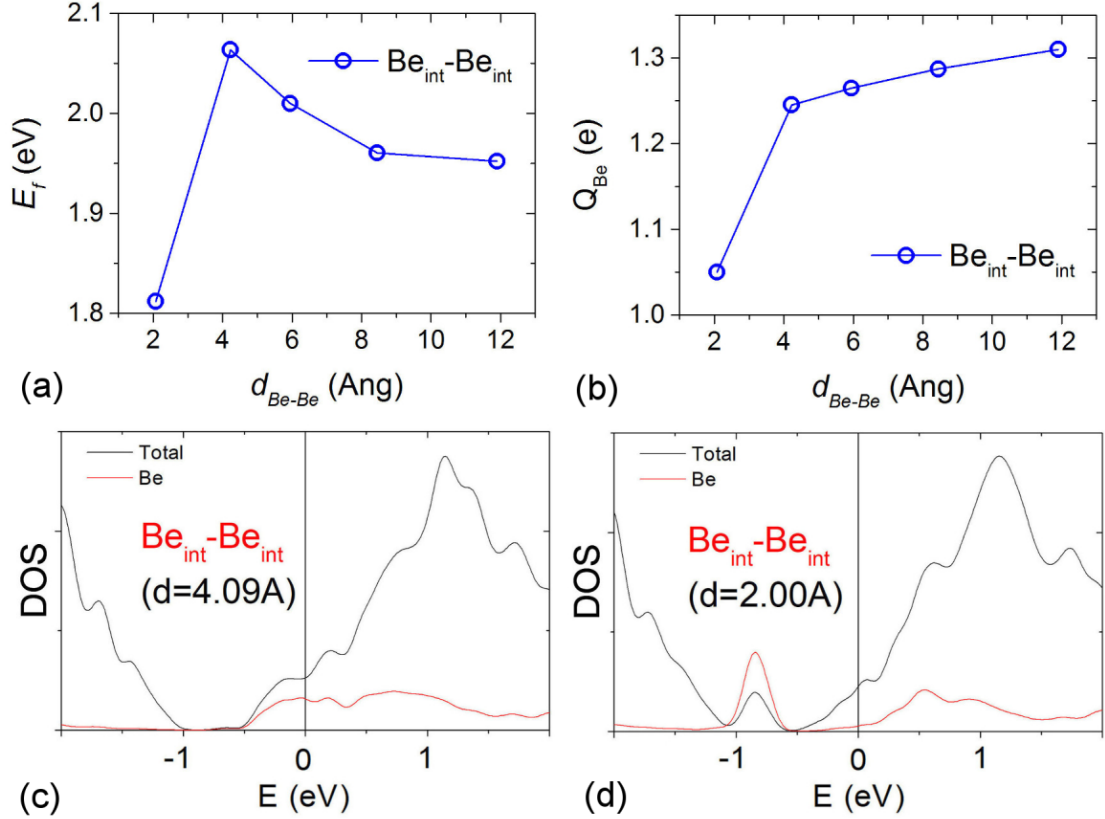

FIG. 3. (a) Formation energies (per Be) and (b) net charges (average per Be) of two Be interstitial dopants versus Be-Be-distance. (c) The density of states of the double $Be_{int}$- $Be_{int}$ defect at the far and (d) near separation. Zero energy is set to the Fermi level. Be contribution to the DOS is magnified.

This behaviour is consistent with that of alkali and alkali-earth atom interstitials reported in other semiconductors[10-15, 18-26]. However, as the two Be interstitial atoms assume nearest-neighbour sites, the defect formation energy per Be drops from 2.07 eV (at a separation of about 4 Å) to 1.81 eV. This corresponds to a significant change in Bader charges from about +1.3 |$e$| at further distances to +1.05 |$e$|. The formation of the Be - Be interstitial pair is accompanied by the kick-out of a Ga atom from its original position into the interstitial space. One of the Be atoms forms a stable 2Be - Ga dumbbell structure shown in Fig. 1(d). The kicked out Ga atom changes its charge from +0.65 to +0.49 |$e$|, see Table II, where Bader charges on atoms involved in this complex are listed.

**TABLE II.** Charges on atoms in the 2Be+Ga dumbbell complex.

| d(Be-Be), Å | Q, |e| |
|---|---|
| Be1 | +1.42 |
| Be2 | +1.34 |
| Ga (complex) | +0.49 |
| Ga (away from complex) | +0.65 |

This corresponds to significant changes in the electronic structure as illustrated in Fig. 3(c,d). The $Be_{int}$ - $Be_{int}$ nearest neighbour pair is metastable: there is a thermodynamic barrier of about 0.26 eV (Fig. 3(a)) for its dissociation, but the defect formation energy is somewhat lower than for well-separated defects.

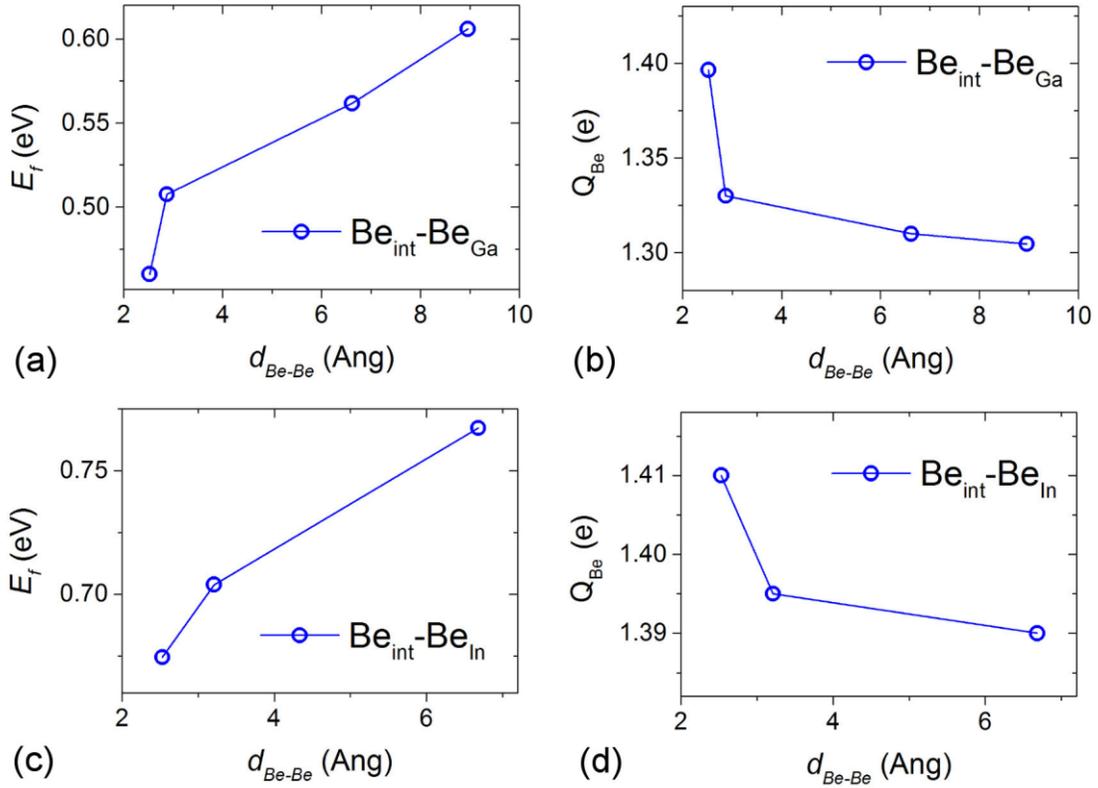

FIG. 4. Defect formation energy per Be and charges on Be atoms of a double substitutional-interstitial (a-b) $Be_{int}$-$Be_{Ga}$ and (c-d) $Be_{int}$-$Be_{In}$ defects.

We note that this barrier is on the order of $3kT$ for typical Be diffusion experiments (where temperatures up to 1000 K are common), we therefore expect this metastable double defect, not yet considered in literature, to play a non-negligible role. It should therefore be included in further more accurate simulations of Be doped InGaAs and specifically Be diffusion.[44]

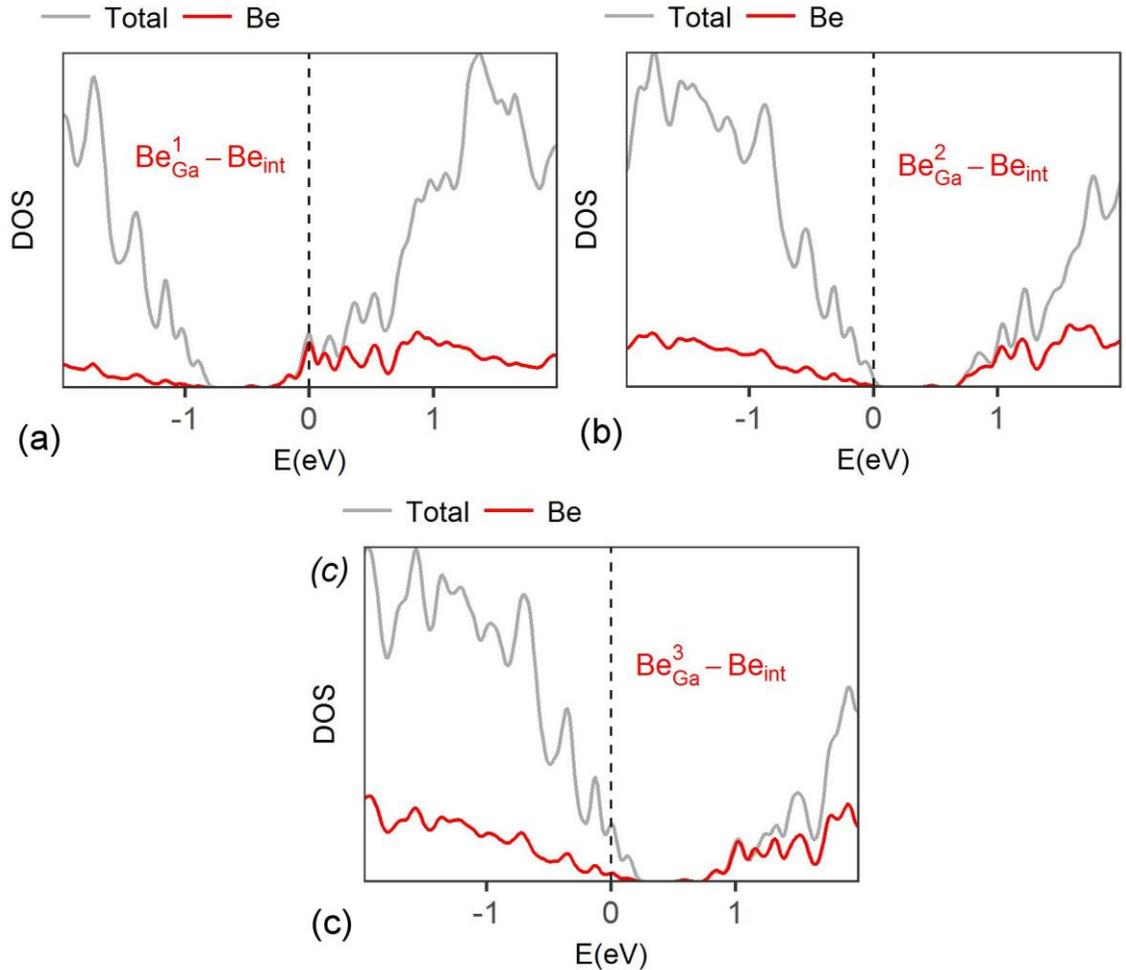

FIG. 5. The density of states of the double $Be_{Ga}$- $Be_{int}$ defect. Zero energy is set to the Fermi level. Be contribution to the DOS is magnified by 15.

We now consider the interactions between interstitial and substitutional Be defects. Non-interacting defects would result in a defect formation energy (per Be atom) which is the average of the numbers listed in Table I. For example, for $Be_{int}$ - $Be_{In}$, one expects $E_f = 1.18$ eV and for $Be_{int}$ - $Be_{Ga}$, 1.06 eV. Fig. 4 shows the defect formation energy of the double $Be_{int}$-$Be_{In}$ and $Be_{int}$-$Be_{Ga}$ defects as a function of the Be - Be distance. Even at the largest Be - Be distance that can be accommodated by the simulation cell, the $E_f$ is significantly stabilized (for example, $E_f$ = 0.61 eV for the $Be_{int}$ − $Be_{Ga}$ double defect). The analysis of the DOS, shown in Fig. 5, reveals the reason for this stabilization: the electrons donated by the interstitial Be fill holes created in the valence band by the substitutional Be atoms instead of occupying higher-energy conduction band states (in the absence of interstitials). This is in contrast to the double interstitial defect considered above where both Be interstitial atoms donate electrons to the conduction band of the host. Specifically, by comparing the DOS of $Be_{Ga}$ - $Be_{int}$ and 2,3^$Be_{Ga}$ - $Be_{int}$ (i.e. a system with 1 substitutional and 2 or 3 substitutional $Be_{Ga}$), it appears that one substitutional $Be_{Ga}$ creates one hole (one unoccupied state) in the valence band (VB); this is expected as Be has one fewer valence electrons than Ga (or In) it replaces. This hole accommodates one of the valence electrons of $Be_{int}$ (the other remaining in the conduction band (CB) as is evidenced by the Fermi level in the CB), and in the presence of 2 such substitutionals, both valence electrons of $Be_{int}$ are accommodated in the VB (as is evidenced by the Fermi level moving to the VB). That it takes two $Be_{Ga}$ substitutional defects for the Femi level to move to the valence band is an indication that Be interstitial donates two electrons to the CB and justifies assignment of the Bader charge of about +1.4 |e| to a 0 K charge state of +2 in the previous subsection. This also proves that one $Be_{Ga}$ (or $Be_{In}$) creates one hole in the VB. This might also justify an assignment of the Bader charge of about +1.4 |e| for interstitial Be to an integer charge state +1 rather than +2, for purposes of Be diffusion modelling. A similar picture holds for the double defect $Be_{In}$ - $Be_{int}$. This is the mechanism similar to that leading to stabilization of the substitutional-interstitial pair of Mg atoms doped into Si or Ge described in Refs. [11, 13]. The difference is that in Si and Ge, a substitutional Mg creates two hole states in the valence band, which was apparent in the DOS analysis performed in Ref. [11] and is explained by a difference of two in the valence between Mg and Si or Ge. As a result, Mg dopants in Ge prefer to distribute equally between substitutional and interstitial sites, while in the present case the preferred ratio (i.e. that minimizing the energy) is 1 : 2 between $Be_{int}$ : $Be_{Ga/In}$. Another way to look at this system is to consider the insertion energy, $E_{ins}$ of $Be_{int}$ into InGaAs doped with substitutional Be.

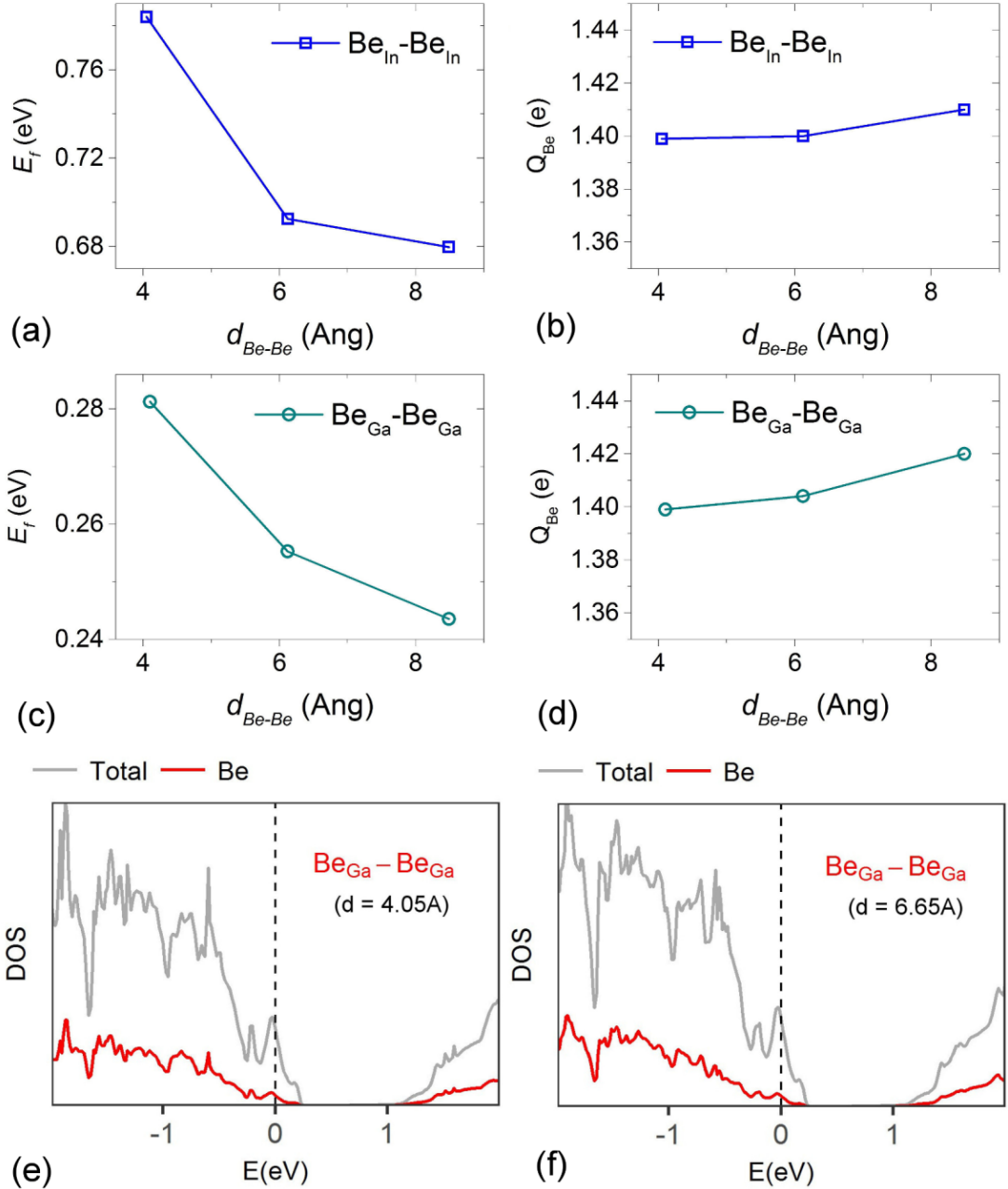

FIG. 6. Formation energies and Bader charges of lower-energy (see Table I) substitutional-substitutional double Be defects: (a-b) $Be_{In}$-$Be_{In}$, (c-d) $Be_{Ga}$-$Be_{Ga}$. (e, f) The density of states of the double $Be_{Ga}$- $Be_{Ga}$ defects at different separations. Zero energy is set to the Fermi level. Be contribution to the DOS is magnified by 15.

While the defect formation energy defined in Eq. (1) is per defect considering all defects present in the simulation cell, the insertion energy $E_{ins}$ only considers the $Be_{int}$ defect formation energy considering substitutionally doped InGaAs as the host; this is given in Table III: the presence of two or more $Be_{Ga}$ substitutionals makes the insertion energy of Be interstitial turn negative, i.e. thermodynamically favourable (compare to Table I).

**TABLE III.** Insertion energy $E_{ins}$ (in eV) of the interstitial $Be_{int}$ defect into InGaAs doped with one, two, and three substitutional $Be_{Ga}$.

| System | $E_{ins}$ |
|---|---|
| 1^$Be_{Ga}$+$Be_{int}$ | 0.852 |
| 2^$Be_{Ga}$+$Be_{int}$ | -0.009 |
| 3^$Be_{Ga}$+$Be_{int}$ | -0.166 |

Contrary to the case of two nearest neighbour interstitial Be, the $Be_{int}$ - $Be_{In}$ and $Be_{int}$ - $Be_{Ga}$ pairs are lower in energy (by about 0.15 eV per Be) than well-separated defects when the two defects are close, with a monotonic increase of $E_f$ with increasing distance (Fig. 4). The interstitial-substitutional pair has a defect formation energy much more competitive with the (lowest-energy) substitutional defects ($Be_{In,Ga}$) than one interstitial Be. Therefore, even though an isolated Be interstitial has a high $E_f$, the presence of substitutional Be defects makes existence of interstitial Be favourable, by the so-called self-doping effect described in Ref. [35]. *This means that interactions between defects in Be doped InGaAs are critically important i.e. they significantly change defect energetics and are expected to significantly influence Be diffusion.*

As for interactions between substitutional Be defects, we find that they are not substantial, as can be seen in Fig. 6 showing distance dependent defect formation energies and Bader charges for the case of $Be_{In}$ - $Be_{In}$ and $Be_{Ga}$ - $Be_{Ga}$ defects. That is, the $E_f$ per Be is close to that of individual defects (Table I) at all distances (mind the scale of Fig. 6), and no changes in the DOS are observed. This is also true for $Be_{Ga}$ - $Be_{In}$. Therefore these interactions could probably be ignored in diffusion models. From Table I it is seen that Ga substitutional (at the In position) defects also possess sufficiently low energy to potentially be present and impact Be diffusion dynamics. $Ga_{int}$ have much lower $E_f$ than As or In interstitials (Table I) and, similar to $Be_{int}$, might be stabilized by substitutional defects, via a similar mechanism. We therefore consider in the following the double defects $Ga_{int}$ - $Ga_{In}$, as well as interactions between Ga and Be defects in the double defects $Be_{in}$ - $Ga_{int}$, $Be_{int}$ - $Ga_{In}$. The results are shown in Fig. 7 and 8 showing the distance-dependent $E_f$ and DOS, respectively. The defect formation energies of $Ga_{int}$ - $Ga_{In}$ are somewhat higher but close to that of non-interacting defects (which is 0.84 eV based on data listed in Table I), i.e. $Ga_{int}$ is not stabilized by $Ga_{In}$. $Be_{int}$ is also not stabilized by $Ga_{In}$. The $Be_{In}$ - $Ga_{int}$ double defect, on the other hand, is significantly stabilized (by about 0.36 eV per defect) vs non-interacting defects by the same $p$-doping mechanism by which $Be_{int}$ is stabilized by Be substitutionals. $Ga_{int}$ only donates one electron into the conduction band, as one $Be_{In}$ is sufficient to bring the Fermi level to the VB, as shown in Fig. 8.

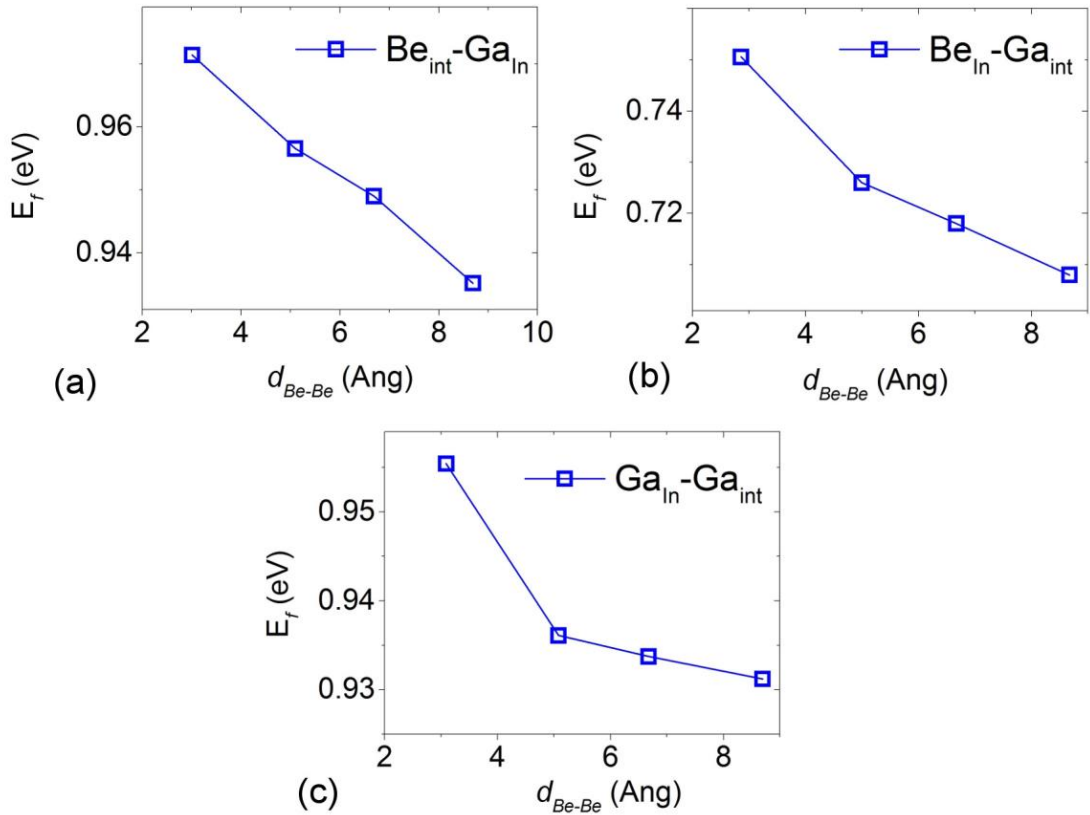

FIG. 7. Formation energies of double defects: (a) $Be_{int}$-$Ga_{In}$, (b) $Be_{In}$-$Ga_{int}$ and (c) $Ga_{In}$-$Ga_{int}$.

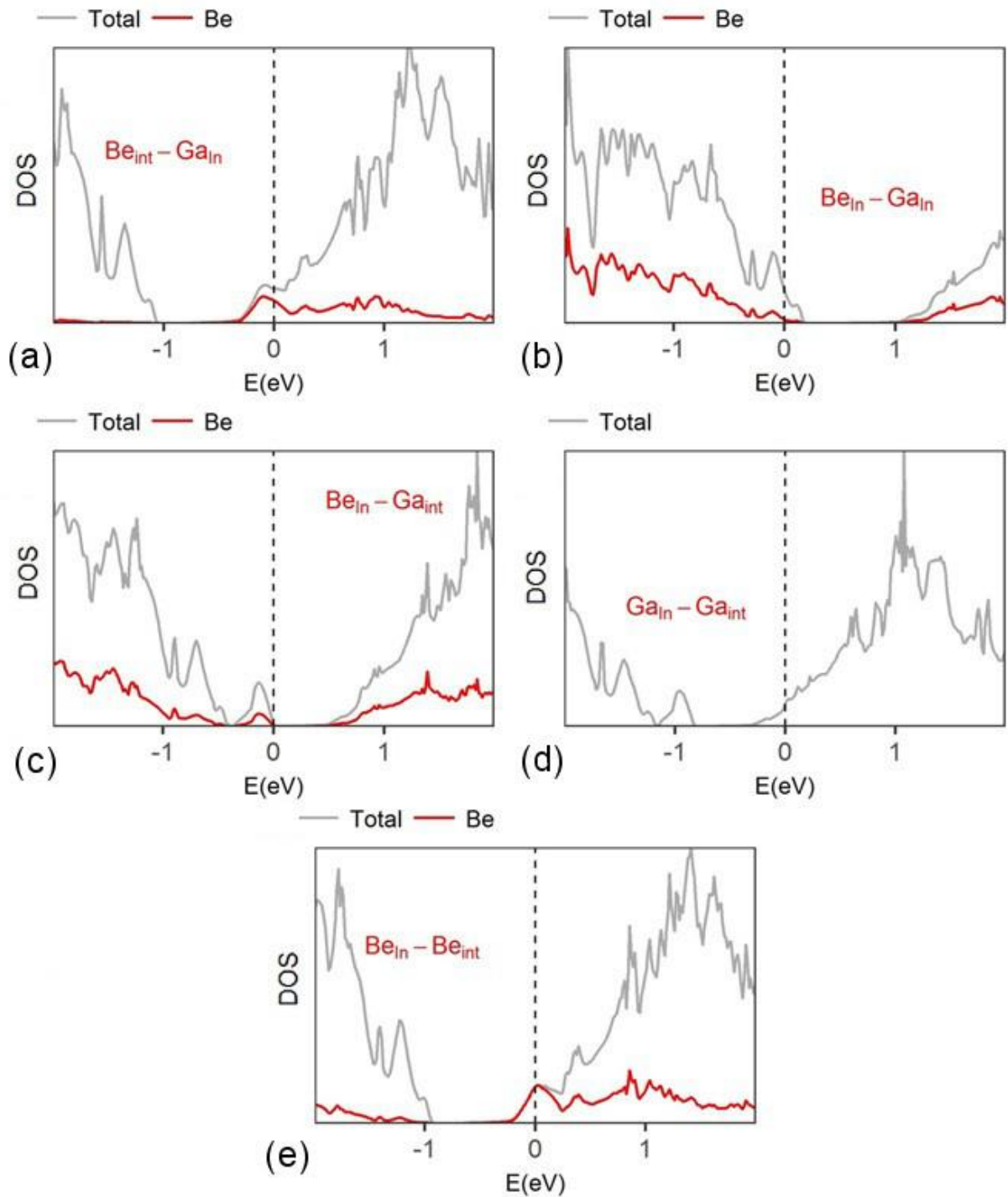

FIG. 8. Density of states of selected double defects. Zero energy is set to the Fermi level. Be contribution to the DOS is magnified by 15.

This corresponds to a Bader charge of only about +0.4 |$e$| Ga$_{int}$, see Table I (compared with +1.4 |$e$| for Be$_{int}$). Ga$_{int}$ therefore may play a role in Be-doped InGaAs. For Ga$_{In}$ - Ga$_{int}$ and Ga$_{In}$ - Be$_{int}$, the Fermi level remains in the conduction band, i.e. Ga$_{In}$ does not seem to create a hole for the electron donated to the CB by these interstitials. This explains why these defects are not stabilized ($E_f$ is slightly higher than for well-separated defects).

## IV. CONCLUSIONS

We presented a computational density functional theory study of interactions between defects in beryllium doped InGaAs. To compare interstitial and substitutional defects, we introduced a model to estimate the energy of substitutional defects in multicomponent solids which is different from the conventional approach in that it permits computing unique, *ab initio* values of defect formation energies in a similar way as it is done in monoelemental hosts. The model essentially treats Be doped InGaAs as a closed system (as opposed to the open system implied by the use of chemical potentials). This is a new perspective and it could be useful for a range of other materials. While substitutional Be (Be$_{Ga,In}$) is much energetically preferred over interstitial (defect formation energy $E_f$ = 0.41/0.18 eV for Be$_{In}$/Be$_{Ga}$ vs 1.95 for Be$_{int}$), we show that changes in electronic structure induced by interstitial (*n*-) and substitutional (*p*-) dopants lead to substantial stabilization of Be$_{int}$ in the presence of substitutional Be, so much so that $E_f$ < 0 for Be$_{int}$ in the presence of 2 Be$_{Ga}$. The mechanism is that of accommodation of electrons, which are donated by interstitial atoms to the conduction band in pure InGaAs, by the lower energy hole states in the valence band created by the substitutional atoms. A similar mechanism also stabilizes Ga$_{int}$ in presence of Be$_{Ga}$/Be$_{In}$. Interstitials therefore should be explicitly considered when modelling Be diffusion in InGaAs. Interactions between two Be$_{int}$ lead to the formation of a metastable Be$_{int}$-Be$_{int}$-Ga defect with a dissociation energy of about 0.26 eV per Be. Our results therefore establish that defect-defect interactions are critically important in Be doped InGaAs and are likely to affect Be diffusion in InGaAs. In contrast, interactions between substitutional defects were found to be small, either due the high energy of or lack of interaction between such defects, and can probably be neglected when modelling Be diffusion, as can high-energy As defects. We suggest that In and Ga should be treated as separate atoms and not lumped into a single effective group III element, as has been done before.[32] In some cases, we identified dopant-centred states which indicate the presence of other charge states at finite temperatures, such as the presence of $Be_{int}^{+1}$ (as opposed to $Be_{int}^{+2}$ at 0K). We hope that these findings will be useful when building accurate models for Be doped of, and Be diffusion in InGaAs.

## V. ACKNOWLEDGEMENTS

This work was supported by the Ministry of Education of Singapore. S.A.C. and W.L. acknowledge industry financial support provided by GlobalFoundries and a research scholarship provided by the Energy Research Institute at Nanyang Technological University (ERIAN). We thank Prof. Siew Ann Cheong, Dr. Mahasin Alam Sk for discussions. We thank Dr. Ignacio Martin-Bragado for discussions, specifically for alerting us to the necessity to more explicitly introduce out approach to defect formation energies and charges considering the existence of abundant literature using chemical potentials and charged simulation cells.